\documentclass[final,5p,times,twocolumn]{elsarticle}

\usepackage{amsmath,amssymb}
\usepackage{comment}
\usepackage{xcolor}
\biboptions{numbers,sort&compress}

\begin{document}

\title{Comment on `Simulation of ultra-relativistic electrons and positrons
channeling in crystals with MBN Explorer'}

\author{Andriy Kostyuk}

\address{Lotzstra{\ss}e 50, 65934 Frankfurt am Main, Germany}
\ead{andriy.p.kostyuk@gmail.com}

\begin{abstract} 
The \textit{snapshot model} of crystal atoms was implemented in 
the Monte Carlo code ChaS (Channeling Simulator) and is being 
successfully used for simulation of ultrarelativistic particle 
channeling. The model was criticized by 
Sushko \textit{et al.} (J. Comp. Phys. 252 (2013) 404--418)
who claim that it overestimates the mean scattering 
angle in a single projectile-atom collision.
As a matter of fact, no evidence that would support this claim
can be found in the mentioned publication.
Moreover, the snapshot model and the model
suggested by Sushko \textit{et al.} 
yield essentially the same value of the 
mean scattering angle. Contrary to the claim of 
Sushko \textit{et al.}, the target electrons 
can be considered 
as fixed-position scatterer, 
corrections due to their finite mass and nonzero initial velocity
have a negligible impact on the channeling 
of light projectiles (electrons and positrons).
In contrast to the snapshot model, the 
model preferred by Sushko \textit{et al.} does not 
take into account incoherent scattering of the projectile
by crystal electrons. This explains why the two models predict
different values of the dechanneling length. 
The claim that the snapshot model underestimates the 
dechanneling length is unfounded. In actual fact, 
this model is in good agreement with experimental data.
\end{abstract}


\maketitle

In Ref. \cite{Sushko2013a}, Sushko \textit{et al.} criticized the \textit{snapshot model} 
of crystal atoms \cite{Kostyuk2010,Kostyuk2011a}.
Instead, they suggested to substitute the atom with its average potential. 
In the following, the model used in Ref. \cite{Sushko2013a}
will be referred to as the average atomic potential (AAP) model.
Sushko \textit{et al.} implemented the AAP model in the code MBN Explorer.

Unlike the AAP model, the snapshot model assumes that atomic electrons are seen by a 
high-speed projectile
as point-like charges at random positions around the nucleus.
The atomic potential is therefore modeled as a sum of   
Coulomb potentials of the nucleus and the electrons (see \cite{Kostyuk2010}
for details).
Having been implemented in the computer code ChaS (Channeling Simulator), this model is
used for Monte Carlo simulation of ultrarelativistic particle channeling in crystals
\cite{Kostyuk2010,Kostyuk2011a,Kostyuk2011c,Kostyuk2012,Kostyuk2013d,Kostyuk2017}.

It is stated in Ref. \cite{Sushko2013a} that the snapshot approximation overestimates
the mean scattering angle in a single projectile-atom collision. 
In actual fact, neither the main paper text nor the supplementary material 
of Ref. \cite{Sushko2013a} 
contain calculation results for the mean scattering angle. 
It is not explained what value was considered as a benchmark and 
how the authors came to the conclusion that the value of 
the mean scattering angle calculated in the snapshot model
overestimates the benchmark.

Moreover, it can be shown that the mean scattering angle
in the snapshot model 
is equal to the the scattering angle in the AAP model
at the same value of the impact parameter unless the models are used 
outside of their applicability domain.
Indeed, the point-like electrons in the snapshot model are distributed around the
nucleus with the probability density 
\begin{equation}
w(\mathbf{r}) = \frac{\nabla^2  U_\mathrm{M}(r)}{4 \pi e Z},
 \end{equation}
where $\mathbf{r}$ is the radius-vector with the  origin in the nucleus,
$e$ is the absolute value of the elementary charge and $Z$ is the atomic number of the 
atom, $U_\mathrm{M}(r)$ is the average atomic potential approximated with
Moli{\`e}re's parametrization \cite{Moliere1947}. (The same approximation for the potential 
is used in the AAP model \cite{Sushko2013a} too.)
Due to its spherical symmetry, Moli{\`e}re's potential
depends only on the absolute value of the radius vector $r = |\mathbf{r}|$.

As it has already been mentioned above, 
the potential $U_\mathrm{S}(\mathbf{r})$ of the \textit{snapshot} atom is defined 
as a sum of the Coulomb potentials of the the nucleus and the point-like electrons.
Due to linearity of Poisson's equation, averaging of $U_\mathrm{S}(\mathbf{r})$
over a large number of snapshots reproduces the original Moli{\`e}re's potential 
\begin{equation}
\langle U_\mathrm{S}(\mathbf{r}) \rangle = U_\mathrm{M}(r) . \label{potentials}
\end{equation}
Sushko \textit{et al.} refer to the above result in Eq. S4 
and check it numerically in Fig. S1 of the 
supplementary material of Ref. \cite{Sushko2013a}.
However, they did not do further steps towards the evaluation of the mean scattering
angle. Let us fill this gap. 

It follows from (\ref{potentials}) that a similar 
expression is valid for the force $\mathbf{F}_\mathrm{S}(\mathrm{r})$ acting on the 
projectile of charge $\mathfrak{q}$ in the field of the  \textit{snapshot} atom 
\begin{equation}
\langle \mathbf{F}_\mathrm{S}(\mathbf{r}) \rangle = \langle - \mathfrak{q} \nabla U_\mathrm{S}(\mathbf{r}) \rangle 
= - \mathfrak{q} \nabla \langle  U_\mathrm{S}(\mathbf{r}) \rangle
= - \mathfrak{q} \nabla U_\mathrm{M}(r) = \mathbf{F}_\mathrm{M}(\mathbf{r}), 
\end{equation}
where $\mathbf{F}_\mathrm{M}(\mathbf{r})$ is the force acting on the same projectile in
Moli{\`e}re's potential $U_\mathrm{M}(r)$. The force is directed along the radius vector,
$\mathbf{F}_\mathrm{M}(\mathbf{r}) = \mathbf{r}/r F_\mathrm{M}(r)$, due 
to spherical symmetry of Moli{\`e}re's potential.

In a similar way, any other quantity that depends linearly on the potential 
has the same value in the both models.

In the case of channeling,
scattering through only very small angles (of the order of Lindhard's critical angle, 
Eq. (1.1) of Ref. \cite{Lindhard1965}) is relevant. Typically, Lindhard's angle is
smaller than one milliradian.
Therefore, the approximation $\sin \theta \approx \theta$ can be
safely used. \label{page_footnote_classical}
The trajectory\footnote{\label{footnote_classical}
The both models considered in the 
present comment are designed for modeling of projectile interaction
with crystal media. Therefore, we do not consider a scattering of the projectile by an isolated 
atom, that cannot be treated classically, but a contribution of a single atom to the projectile 
interaction with the crystal.
Due to interaction with a large number of atoms, the motion of projectile becomes classical
(see e.g. the discussion on p. 3 of Ref. \cite{Kostyuk2010}) provided that its energy is sufficiently
large ($E \gtrsim 100$ MeV for electrons and positrons). 
This justifies using the notion of trajectory in the present calculations.}
of the projectile during the scattering 
can be approximated with a straight line along the initial direction of particle momentum.
This allows one to express the scattering angle 
$\pmb{\theta}_\mathrm{S}(\mathbf{b})$
as a linear functional of the force and to show that its average over snapshots
is equal to the scattering angle 
$\pmb{\theta}_\mathrm{M}(\mathbf{b})$ in Moli{\`e}re's potential:
\begin{eqnarray}
\langle \pmb{\theta}_\mathrm{S}(\mathbf{b})  \rangle &=&
\left \langle \frac{1}{v p}  \int_{-\infty}^{+\infty}  \mathbf{F}_{\mathrm{S} \perp}(\mathbf{b},r_{\parallel}) dr_{\parallel}   \right \rangle  
\nonumber \\
&=& \frac{1}{v p}  \int_{-\infty}^{+\infty}  \mathbf{F}_{\mathrm{M} \perp}(\mathbf{b},r_{\parallel}) dr_{\parallel}  =
\pmb{\theta}_\mathrm{M}(\mathbf{b}),
\label{mean_scattering_angle}
\end{eqnarray}
where $v$ and $p$ are, respectively, projectile speed and momentum, subscripts $\parallel$ and $\perp$ stand, respectively, 
for longitudinal and transverse components of the corresponding vectors with respect to the initial direction of the projectile
momentum and $\mathbf{b}$ is a vector connecting the projections of the atomic nucleus and the initial position of the projectile
onto the transverse plane. The absolute value $b$ of the transverse vector $\mathbf{b}$ is the impact parameter of the collision.
The scattering angle $\pmb{\theta}$ is also a transverse vector. In a spherically-symmetric potential, it has only 
a radial component $\pmb{\theta}_\mathrm{M}(\mathbf{b}) = \mathbf{b}/b \theta_\mathrm{M}(b)$. The same is true for the 
mean scattering angle in the snapshot model,
$\langle \pmb{\theta}_\mathrm{S}(\mathbf{b})  \rangle = \mathbf{b}/b \langle \theta_\mathrm{S} \rangle(b) $, although the 
transverse vector 
$\pmb{\theta}_\mathrm{S}(\mathbf{b})$ in a single snapshot may have a nonzero azimuthal component.

So, the mean scattering angle in the snapshot model is equal to the the scattering angle in AAP model
unless it is large enough to invalidate the approximations that have been used to obtain 
Eq. (\ref{mean_scattering_angle}),
i.e. unless it exceeds Lindhard's angle by orders of magnitude. If such a large angle scattering takes place, only 
the fact that the projectile quits the channelling or quasi-channelling mode is essential. The precise direction of the projectile 
motion after a large angle scattering is irrelevant to the channeling process and does not influence the 
simulation results.

Instead of evaluating the mean scattering angle, as it is announced in the main text of the paper, 
Sushko \textit{et al.} introduced and evaluated another quantity --- the root mean square (rms) 
scattering angle
\begin{equation}
\bar{\theta}(b) = \sqrt{\langle \pmb{\theta}^{2}(b) \rangle} \label{rms_theta}
\end{equation}
in the supplementary material of Ref. \cite{Sushko2013a}.
Unlike the mean scattering angle, the rms scattering angle is a nonlinear 
functional of the force. For this reason, its values in the two models
are different.

The authors of Ref. \cite{Sushko2013a} evaluated numerically\footnote{The 
numerical procedure is based on Eq. (S1) of the supplementary material of  
Ref. \cite{Sushko2013a}. The applicability domain of this formula is limited
to small scattering angles.  Sushko \textit{et al.} mention this fact on 
page S1, but they do not reveal if this restriction is enforced 
in their numerical procedure.
If not properly used, Eq. (S1) can yield arbitrary large values
of the scattering angle that destroy the convergence of the calculation.
This may be the reason for the large statistical errors seen in Fig. S2
and for their irregular dependence on the impact parameter.}
the rms 
scattering angle in the snapshot model $\bar{\theta}_\mathrm{S}(b)$ 
and plotted it in Fig. S2 of the supplementary material. The calculated 
value of $\bar{\theta}_\mathrm{S}(b)$ 
is noticeably larger than the corresponding 
value for the AAP model if the impact parameter exceeds the 
Thomas-Fermi radius, $b>a_\mathrm{TF}$.
Sushko \textit{et al.} attribute this difference to  
the fact that target electrons are treated by the snapshot 
model as motionless infinitely heavy and, therefore, 
\textit{`the recoil of the scatterer is fully ignored in the collisional
process'}. 
This statement cannot be valid, because 
the algorithm of MBN Explorer, similarly to the one of ChaS, takes into account neither
the finite mass of the target electrons nor the recoil of the scatterer.\footnote{
Authors of  Ref. \cite{Sushko2013a}
do consider another recoil effect, the radiative 
recoil, which causes projectile energy loss due to photon emission. 
However, this effect is not related to  the recoil of the scatterer.
Moreover, 
in view of the fact that emission of hard photons by channelled moderate-energy,
$E \lesssim 1$ GeV, electrons and positrons is a rare event, its influence 
on the channeling process is negligible.} Therefore, the mentioned effects 
cannot cause the difference between the results of two models.

Nonetheless, let us see how strong is the impact of the finite mass of the 
target electron and its nonzero initial velocity  on small-angle scattering 
of light projectiles: electrons and positrons. First, let us compare 
the  differential cross section of positron/electron scattering 
by a target electron within the approximation used in the code ChaS to 
M{\o}ller's and Bhabha's cross sections for small scattering 
angles.

From 
Eq.(2) of Ref. \cite{Kostyuk2017a} one obtains 
the transverse momentum gained by an ultrarelativistic electron or positron
due to collision with a target electron 
\begin{equation}
\Delta \vec{p}_{\perp} = \eta \frac{2 e^2}{v b^2} \vec{b},  
\label{pperp}
\end{equation}
where $\eta = \pm 1$  stands for the sign of the projectile electric charge,
$e$ is the elementary charge, $v$ is the projectile speed,  $\vec{b}$
is the vector connecting the projections of the initial positions of the 
projectile and the target onto transverse plane, $b=|\vec{b}|$ is the impact
parameter of the collision.
Then the scattering angle $\theta(b)$ satisfies the following relation
\begin{equation}
\sin \theta(b) =  \frac{| \Delta \vec{p}_{\perp} (\vec{b}) |}{P}   =  \frac{2 e^2}{v b P}, 
\end{equation}
where  $P$ is the absolute value of the projectile
momentum, which is assumed to be the same before and after the collision 
(the loss of the projectile energy is neglected).
Taking into account $d \sigma = 2 \pi b \, d b$ one obtains the differential 
cross section
\begin{equation}
\frac{d \sigma_\mathrm{ChaS}}{d \Omega}  =  
\frac{4 r_\mathrm{e}^2}{\gamma^2 \beta^4} 
\frac{\cos \theta}{\sin^4 \! \theta} ,
\label{xsection_ChaS}
\end{equation}
where $r_\mathrm{e}$ is the classical electron radius, $\beta = v / c \approx 1$ is  
the ratio of the projectile speed to the speed of light and 
$\gamma = 1/\sqrt{1 - \beta^2}$ is the Lorentz factor of the projectile.

Let us express M{\o}ller's and Bhabha's cross sections 
(Eqs.(81.7) and (81.17) of Ref.\cite{Berestetskii1982}, respectively)
in terms of the scattering angle $\theta$ of one of the particles in the 
initial rest frame of the other one and expand the result in powers of 
$\sin \theta$:
\begin{eqnarray}
\frac{d \sigma_\mathrm{Moller}}{d \Omega } \! \! & = & \! \! \frac{4 r_\mathrm{e}^2}{\gamma^2 \beta^4}
\frac{\cos\theta}{\sin^4\!\theta} 
\left [ 1- \kappa_\mathrm{M} \sin^2\!\theta  
+ O\left(\gamma^2 \sin^4\!\theta\right)
\right ] \label{xsection_Moller} \\
\kappa_\mathrm{M} \! \!  & = & \! \!  1 + \frac{1}{2 \gamma} - \frac{1}{2 \gamma^2}   \approx 1 \mbox{ for } \gamma \gg 1
\label{kappa_Moller} \\
& & \nonumber \\
\frac{d \sigma_\mathrm{Bhabha}}{d \Omega} \! \! & = & \! \! \frac{4  r_\mathrm{e}^2}{\gamma^2 \beta^4} \frac{\cos\theta}{\sin^4\!\theta}
\left[
1 +  \kappa_\mathrm{B} \sin^2\!\theta + O\left( \gamma^2 \sin^4\!\theta\right) \right ] 
\label{xsection_Bhabha} \\   
\kappa_\mathrm{B} \! \!  & = & \! \!  \gamma  \left(1 - \frac{1}{2 \gamma^2}\right) \left(1 - \frac{1}{\gamma}\right)  \approx \gamma \mbox{ for } \gamma \gg 1
\label{kappa_Bhabha}
\end{eqnarray}

The leading term is the same in Eqs. (\ref{xsection_ChaS}), (\ref{xsection_Moller}) and (\ref{xsection_Bhabha}). 
Therefore, the difference between the three formulas has to be negligible if the scattering angle $\theta$ is small enough.
Let us estimate the relative errors of Eq. (\ref{xsection_ChaS}) with respect to (\ref{xsection_Moller}) and 
(\ref{xsection_Bhabha}) at $\theta=\theta_{\mathrm{L}}$, where $\theta_{\mathrm{L}}$ is Lindhard's critical angle \cite{Lindhard1965}: 
\begin{equation}
 \theta_\mathrm{L} \approx \sin \theta_\mathrm{L} = \sqrt{\frac{2 U_{\max} E}{P^2 c^2}} = \sqrt{\frac{2 U_{\max}}{\beta^2 \gamma m_\mathrm{e} c^2}},
\end{equation}
where $m_\mathrm{e}$ is the electron mass, $m_\mathrm{e} c^2 \approx 500$ keV, and 
$U_{\max}$ is the depth of the planar channel that ranges from a few eV to a few tens of eV, e.g. $U_{\max} \approx 23$ eV for (110) planar channel in silicon.
The relative error of Eq. (\ref{xsection_ChaS}) in estimation of Bhabha's cross section (\ref{xsection_Bhabha}), i.e. for positron projectile, is
\begin{equation}
\kappa_\mathrm{B}  \sin^2\!\theta_{\mathrm{L}} \approx  \gamma \theta_{\mathrm{L}}^2 = \frac{2 U_{\max}}{\beta^2  m_\mathrm{e} c^2} \sim 10^{-4}
\label{error_Bhabha}
\end{equation}
For M{\o}ller's cross section the corresponding quantity is even smaller 
\begin{equation}
\kappa_\mathrm{M}  \sin^2\!\theta_{\mathrm{L}} \approx  \theta_{\mathrm{L}}^2  \sim \frac{10^{-4}}{\gamma}
\label{error_Moller}
\end{equation}
Therefore, the finite mass of the target electron (as well as all other effects that are present in  M{\o}ller's and Bhabha's cross sections and
are neglected in Eq. (\ref{xsection_ChaS})) becomes important if the scattering angle exceeds Lindhard's angle $\theta_{\mathrm{L}}$
by orders of magnitude. As it was already explained above, inaccuracies in the determination of large scattering angles 
do not influence the simulation results. 

It is not difficult to replace Eq.(2) of Ref. \cite{Kostyuk2017a} with a numerical procedure that 
reproduces the scattering cross section exactly.
Such replacement may be necessary if channeling of heavy projectiles (e.g. protons or heavy ions)
has to be modeled. In the case of electron and positron channeling it would only waste the computer
time without any influence on the simulation results. Therefore, a simple model of scattering,
Eq.(2) of Ref. \cite{Kostyuk2017a}, is, in fact, an advantage of the code ChaS, not its drawback.

One more effect of the finite mass of the target electron which is neglected by the algorithm
of ChaS as well as by MBN Explorer are losses of the projectile energy and longitudinal 
momentum due to ionization and excitation of target atoms.

The longitudinal momentum transfer from the 
projectile to a free target particle can be related to the transversal momentum transfer due to
energy-momentum conservation.  
The longitudinal momentum of a light ultrarelativistic projectile (electron 
or positron) is changed in a collision with a free electron by
\begin{equation}
\Delta p_\parallel 
 =  
 - \frac{\gamma +  1}{2 P} \left( \Delta \vec{p}_\perp \right)^2 
\left[ 1 + O \left( \frac{\left( \Delta \vec{p}_\perp \right)^2}{P^2} \right) \right]
 \approx
 - \frac{\gamma}{2} \theta^2  P .
\label{pparallel}
\end{equation}
The relative longitudinal momentum loss at $\theta=\theta_{\mathrm{L}}$ can be estimated 
similarly to Eq. (\ref{error_Bhabha}). 
It amounts to about $10^{-4}$. For projectile energy of the order of $1$ GeV, it 
results into the energy transfer
of about $100$ keV.  This exceeds the binding energy of the atomic electrons by 2 to 5 orders 
of magnitude, which justifies the consideration of the target electron as a free particle.
Although the transferred energy and longitudinal 
momentum have a substantial impact on the target atom, they are very small
comparing to the energy and longitudinal momentum
of the projectile. Therefore, their effect on the 
projectile motion can be neglected. Moreover, projectile scattering through
Lindhard's angle $\theta_\mathrm{L}$ in a collision with a target electron 
is a rather rare event during the channeling process. Most of collisions result into 
much smaller scattering angle, and, consequently, into a much smaller projectile 
energy loss.

In the above consideration, the target electron was in rest
in the lab frame. In reality, the momentum of a bound electron in an atom is not zero
due to Heisenberg's uncertainty principle.
Let us see how strong is the influence of the target electron motion on the 
scattering of the projectile.

Suppose the target electron moves relative to the lab frame in a transversal direction 
with respect to the initial projectile momentum direction with the velocity $\vec{v}_1$.
Using Eqs. (\ref{pperp}) and (\ref{pparallel}) in the rest frame of the target electron and 
and then performing Lorentz's transformation to the lab frame one obtains the following 
expression for the transverse momentum gained by the projectile in the collision
\begin{eqnarray}
\Delta \vec{p}_{\perp}(\vec{\beta}_1)
& = & 
 \Delta \vec{p}_{\perp} \left( 1 -  \frac{\vec{\beta}_1^2}{2 \gamma^2}  
 + \frac{(\vec{\beta}_1 \cdot \vec{b})^2}{b^2} \frac{1}{\gamma^2} \right)
\label{pperpbeta1} \\ 
& & 
-   \frac{\vec{\beta}_1}{\gamma^2} (\Delta \vec{p}_{\perp} \cdot \vec{\beta}_1) 
 +  \frac{\left( \Delta \vec{p}_{\perp} \right)^2}{P} \frac{\vec{\beta}_1}{2} 
 +  \dots  , \nonumber
\end{eqnarray}
where $\vec{p}_{\perp}$ in the right-hand side is given by Eq. (\ref{pperp}),
$\vec{\beta}_1 = \vec{v}_1/c$ and the ellipsis stand for higher order terms 
with respect to $1/\gamma$ and $|\Delta \vec{p}_{\perp}|/P$.

Neglecting the last term in Eq. (\ref{pperpbeta1}) results into a relative 
error of 
\begin{equation}
\frac{| \Delta \vec{p}_{\perp} |}{P} \frac{|\vec{\beta}_1|}{2}  = \theta \frac{|\vec{\beta}_1|}{2},
\end{equation}
which is negligible if scattering angle $\theta$ is sufficiently small. In particular, 
at $\theta = \theta_\mathrm{L}$ the relative error is about 
$10^{-2} \cdot |\vec{\beta}_1| / \sqrt{\gamma}$.
Dropping other terms that contain $\vec{\beta}_1$ leads to a relative error of the 
order of $\vec{\beta}_1^2/\gamma^2$ which is small provided that 
$\gamma \gg 1$ and does not depend on the scattering angle.

Negligibility of the longitudinal motion of the target electron 
is obvious without explicit calculation. 
Provided that the target electron is nonrelativistic
in the lab frame, the Lorentz factor of the projectile in the rest frame of the 
target lectron is large, $\gamma' \gg 1$,  as long as it is large in the 
lab frame, $\gamma \gg 1$. The gained transverse momentum (\ref{pperp}) of the 
projectile depends on the Lorentz factor through the projectile speed in the rest 
frame of the target electron
\begin{equation}
 v' = c \beta' = c \sqrt{1 - \frac{1}{\gamma'^2}} 
 \approx c \left( 1 - \frac{1}{2 \gamma'^2} \right)
\end{equation}
Neglecting the projectile motion is the lab frame means replacing 
$\gamma'$ with $\gamma$. This leads to the relative error of the 
order of 
\begin{equation}
\left | \frac{1}{2 \gamma'^2} - \frac{1}{2 \gamma^2} \right |
< \frac{1}{2 \gamma^2}.
\end{equation}

The above consideration quantifies the qualitative justification
of the snapshot model given in  Ref. \cite{Kostyuk2010}.
It confirms that 
the motion of the electrons of crystal atoms can be neglected if channeling
of ultrarelativistic projectile is studied. 
It also refutes the statements of Ref. \cite{Bezchastnov2018} that 
the atomic electrons 
\textit{`by no means can be considered as static charges'},
the snapshot model \textit{`cannot consistently account for the
electron momenta in the collision process with the projectile'}
and that \textit{`the model introduces a non-controllable
uncertainty in the scattering angle in each individual 
scattering event'}. In fact, the target electron can be considered
as static charges as long as small-angle scattering of light 
projectiles (electrons and positrons) are concerned.
The errors introduced by neglecting 
the momenta of target electrons are well below the per mil level 
and, therefore, are well controllable.

We have thus seen that neither the recoil of scatterer nor 
the motion of atomic electrons affect the scattering of the 
projectile substantially. 
Neglecting these effects cannot not cause any significant 
differences in the model results.
To elucidate the real reason for the difference between the snapshot model 
and the AAP model, let us apply these two approaches to scattering of a fast 
electron (or positron) by an isolated atom. In this case, in contrast to 
the modeling of particle channeling,
the projectile cannot be treated 
classically. To facilitate comparing of the results with textbook formulas, we shall 
consider scattering cross section of fast but nonrelativistic projectile 
using Born's approximation
(see e.g. Eq. (126.7) of Ref. \cite{Landau1991}):
\begin{equation}
\frac{d \sigma}{d \Omega} =
\frac{m^2}{4 \pi^2 \hbar^4}
\left | \int U (\vec{r}) \exp(- i \vec{q} \cdot \vec{r}) \, d^3 \! r \right |^2 ,
\label{Born}
\end{equation}
where $m$ is the mass of projectile,
$m=m_\mathrm{e}$ in our case, and $\hbar$ is Planck's constant.
The vector $\vec{q}$ in Eq. (\ref{Born}) is proportional to the momentum transfer:
\begin{equation}
\vec{q} = \frac{\vec{P}'-\vec{P}}{\hbar},
\end{equation}
where $\vec{P}$ and $\vec{P}'$ are initial and final momentum of the projectile.  

Substituting the potential of a `snapshot' of the target atom,
\begin{equation}
 U_{\{\vec{r}_{j}\}} (\vec{r}) = - \frac{e^2 Z}{r} + \sum_{j=1}^{Z} \frac{e^2}{|\vec{r}-\vec{r}_{j}|},
 \label{u_snapshot}
\end{equation}
into Eq.(\ref{Born}) we obtain the corresponding scattering cross section:
\begin{eqnarray}
\frac{d \sigma_{\{\vec{r}_{j}\}}}{d \Omega}  & = & 
\frac{4 m_\mathrm{e}^2 e^4}{\hbar^4 q^4} \left \{
Z(Z+1) 
 - 2 Z \sum_{j=1}^{Z} \cos ( \vec{q} \cdot \vec{r}_{j} )  \right. \nonumber \\
& & \left. +
\sum_{k \not = j}
 \exp \left[   i \vec{q} \cdot (\vec{r}_{k} - \vec{r}_{j} ) \right]
\right \}
\end{eqnarray}
Averaging it over a large number of snapshots yields 
\begin{eqnarray}
\frac{d \sigma_\mathrm{S}}{d \Omega}  & = &
\left \langle \frac{d \sigma_{\{\vec{r}_{j}\}}}{d \Omega} \right \rangle =
\frac{4 m_\mathrm{e}^2 e^4}{\hbar^4 q^4} \Bigg \{
Z(Z+1) 
 - 2 Z \, \Re\left[F(q)\right]   \nonumber \\
 & & 
 + \left \langle
\sum_{k \not = j}
 \exp \left[   i \vec{q} \cdot (\vec{r}_{k} - \vec{r}_{j} ) \right]
\right \rangle \Bigg  \},
\label{sigma_average}
\end{eqnarray}
where $\Re\left[F(q)\right]$ is the real part of the atomic form 
factor\footnote{We consider unpolarized atoms therefore an average 
over snapshots depends on $q = | \vec{q} |$  rather then on $\vec{q}$.}
\begin{equation}
 F(q) = \left \langle \sum_{j=1}^{Z} 
 \exp ( - i \vec{q} \cdot \vec{r}_{j} )
\right \rangle . \label{FormFactor}
\end{equation}

Neglecting the projectile energy loss, $|\vec{P}'| \approx |\vec{P}|$, 
one can express $q$  as 
\begin{equation}
q = \frac{2 P}{\hbar} \sin \left(  \frac{\theta}{2} \right)  \approx
\frac{P \theta}{\hbar} = \frac{m_\mathrm{e} v \theta}{\hbar},
\label{q_approx}
\end{equation}
where $v = P/m_\mathrm{e}$ is the projectile speed.
Then Eq. (\ref{sigma_average}) can be presented as 
\begin{equation}
\frac{d \sigma_\mathrm{S}}{d \Omega}  = 
\frac{d \sigma_\mathrm{el}}{d \Omega} +
\frac{d \sigma_\mathrm{ie}}{d \Omega} ,
\label{sigma_sum}
\end{equation}
where 
\begin{equation}
\frac{d \sigma_\mathrm{el}}{d \Omega} =
\left( \frac{2 e^2}{m_\mathrm{e} v^2} \right)^2 | Z -   F(q) |^2  \frac{1}{\vartheta^4} 
\label{sigma_elastic}
\end{equation}
and
\begin{equation}
\frac{d \sigma_\mathrm{ie}}{d \Omega} =
\left( \frac{2 e^2}{m v^2} \right)^2
\left \{
Z - | F(q) |^2  +
 \left \langle
\sum_{k \not = j}
 \exp \left[   i \vec{q} \cdot (\vec{r}_{k} - \vec{r}_{j} ) \right]
\right \rangle \right  \} \frac{1}{\vartheta^4}
\label{sigma_inelastic}
\end{equation}
are, respectively, the differential cross sections of elastic and 
inelastic scattering of fast electrons (or positrons) by a neutral 
atom (cf. Eqs. (139.4) and (148.23) of Ref.\cite{Landau1991}).

One might think that the snapshot model is in contradiction to
Heisenberg's uncertainty principle, because the positions of atomic electrons 
are fixed in each snapshot. The above consideration demonstrates 
that there is no any contradiction. In fact, the model simulates the quantum
uncertainty due to different positioning of electrons in different 
snapshots. Therefore, the snapshot model is able to reproduce the textbook quantum mechanical cross 
section,  Eqs. (\ref{sigma_sum}--\ref{sigma_inelastic}), provided that the 
projectile is treated within the quantum approach. 
It has been shown above that the momentum of the target electron has negligible 
influence on the scattering angle. This justifies neglecting the 
uncertainty of the momentum.

Let us apply the same approach to the AAP model 
used in the MBN Explorer. Inserting the \textit{average} atomic potential, 
\begin{equation}
\left \langle U_{\{\vec{r}_{j}\}} (\vec{r}) \right \rangle
= - \frac{e^2 Z}{r} + \left \langle \sum_{j=1}^{Z} \frac{e^2}{|\vec{r}-\vec{r}_{j}|}
\right \rangle ,
\label{u_average}
\end{equation}
into Eq. (\ref{Born}) one obtains after some trivial algebraic transformations 
the elastic scattering cross section (\ref{sigma_elastic}):
\begin{equation}
\frac{d \sigma_{\langle U \rangle}}{d \Omega} = \frac{d \sigma_\mathrm{el}}{d \Omega}
\label{sigma_MBN}
\end{equation}
in contrast to the sum of elastic and inelastic cross sections,  
Eq. (\ref{sigma_sum}), that was obtained using the snapshot model of the target atom.

Born's perturbative approximation cannot be applied to particle channeling in 
a crystal. Therefore, the both codes, ChaS and MBN Explorer, use a different
approximation to describe the projectile  motion. They treat it classically 
(see also footnote \ref{footnote_classical} on page \pageref{footnote_classical}).
However, the models of the target atoms implemented in the codes are exactly 
the same as in the above quantum mechanical consideration:  
Eqs.(\ref{u_snapshot}) and (\ref{u_average}) for ChaS and MBN Explorer, 
respectively.
Therefore, we conclude from Eqs. (\ref{sigma_sum}) and (\ref{sigma_MBN})
that ChaS takes into account elastic as well as inelastic interactions
of the projectile with the crystal atoms. In contrast, MBN Explorer
takes into account only elastic interactions and ignores inelastic ones.
Clearly, inelastic processes, excitation and ionization of crystal atoms,
do take place when an ultrarelativistic charged projectile penetrates through
the crystal. Therefore, the AAP model of MBN Explorer, which
does not take them into account, is incomplete.

A comparison of the two models allows one to gain a deeper insight on the
way how the inelastic processes influence the projectile motion in a
crystal. The most obvious effect, the loss of the projectile 
energy due to inelastic scattering, 
is, in fact, very small and can be neglected. 
Nonetheless the inelastic processes do affect
the motion of a channeled particle in a less straightforward way.

In the AAP model of MBN Explorer
the scattering centers, the nuclei, are present only in the vicinity of the crystal
planes. The space between the planes is assumed to be completely empty
with the effect of the crystal electrons limited to screening of the Coulomb field 
of the nuclei. In reality, the space between the crystal planes is filled 
with electrons. The electrons can scatter the projectile. 
This effect is obviously ignored in the MBN Explorer, 
but it is taken into account by the snapshot model of the code ChaS.

The relation of the incoherent projectile scattering by the crystal
electrons to the inelastic processes is obvious. The energy-momentum 
transfer from the projectile to the target electron leads to excitation
or ionization of the atom. As it has been demonstrated above, 
the transferred energy and, consequently, the transferred longitudinal
momentum are very small comparing to the energy and longitudinal momentum
of the  ultra-relativistic projectile. Therefore, they can be neglected as long as 
the motion of projectile is studied.
In contrast, the transversal momentum of a channeled projectile is small. 
For this reason, the transversal momentum transfer to the target electrons 
due to inelastic collisions is not negligible and it does 
affect the projectile motion substantially. 

It has to be  emphasized that the statement of the last paragraph does 
not contradict to the result of Eqs. (\ref{error_Bhabha})
and (\ref{error_Moller}) about the negligible effect of the target electron 
mass in the channeling process. The total transversal momentum transfer
to the target electron corresponds to the leading 
term in Eqs. (\ref{xsection_Moller}) and (\ref{xsection_Bhabha}). The motion 
of the hit target electron due to the collision changes the momentum
transfer by a tiny amount with respect to the value it would have in 
the case of fixed or `infinite mass' target charge. 
This modification is contained in the small higher-orders 
terms of Eqs. (\ref{xsection_Moller}) and (\ref{xsection_Bhabha})
that are neglected in the numerical procedure of the code ChaS, c.f.
Eq. (\ref{xsection_ChaS}). 

It is noteworthy that the MBN Explorer, although it uses the 
AAP model for the crystal electrons, actually adopts the snapshot model for 
the description of crystal nuclei. Indeed, the target nuclei are modeled 
as point-like objects randomly placed around their average positions in the 
crystal lattice (see Eq. (6) of Ref. \cite{Sushko2013a}), exactly in the same 
way as it is done in the code ChaS (cf. Eq.(23) of Ref. \cite{Kostyuk2010}). 
This is why the incoherent scattering due to inelastic interaction with 
the lattice nuclei (excitation of phonons) is present in the MBN Explorer 
as well as in the code ChaS. Adopting the average potential approach for 
nuclei in the same way as it was done in the AAP model 
for electrons would mean replacing the point-like nuclei with their 
potentials averaged over the random positions of the nuclei. 
In this case, no incoherent scattering 
of the projectile would take place and the dechanneling length calculated 
with this model would be infinite.

\mbox{}

Having clarified the difference between the two models from the 
physical point of view, let us have a mathematical insight.
The scattering in average potential is deterministic.
The angle of projectile scattering by an atom is uniquely determined 
by the impact parameter. Therefore, 
$\bar{\theta}_\mathrm{M}(b) =   \theta_\mathrm{M}(b) $.

In contrast, the snapshot model takes into account incoherent 
scattering of the projectile by atomic electrons. Therefore, 
the scattering angle changes randomly from one snapshot to another.  
Let us represent it as
\begin{equation}
\pmb{\theta}_\mathrm{S}(b) = 
\frac{\mathbf{b}}{b} \langle \theta_\mathrm{S} \rangle (b)  + 
\pmb{\delta} (\mathbf{b}),  \label{theta_S_delta}
\end{equation}
where $\pmb{\delta} (\mathbf{b})$ is the deviation of the scattering angle 
in a given snapshot from its average value. By definition, the 
mean value of $\pmb{\delta} (b)$ over a large number of snapshots
is zero:
$\langle \pmb{\delta} (\mathbf{b}) \rangle = \mathbf{0}$. Taking this 
into account one obtains from (\ref{rms_theta}) and (\ref{theta_S_delta})
\begin{equation}
\bar{\theta}_\mathrm{S}(b) 
 =  \sqrt{\langle \theta_\mathrm{S} \rangle^{2} (b) + 
\langle \pmb{\delta} \rangle^{2} (b) } 
=  \sqrt{\theta_\mathrm{M}^{2}(b) + 
\langle \pmb{\delta} \rangle^{2} (b) } \, > 
 \theta_\mathrm{M} (b)  ,
\end{equation}
which is just a well known mathematical fact that the rms value of any set of 
real numbers is always larger than or equals to the magnitude of theirs mean 
value. The equality is 
reached in the only case of all numbers in the set equal each other. 
Therefore, AAP model is a limiting case yielding the least possible 
rms scattering angle for each value of mean scattering angle.

Any model that takes into account incoherent scattering of the projectile
by atomic electrons has to be stochastic: the value of the scattering angle
has to vary randomly even if the impact parameter is fixed. Therefore, not 
only the snapshot model but also any other model that takes into account the 
incoherent scattering has to yield a larger value of the rms scattering 
angle than the AAP model does.

For the same reason, the dechanneling length in the snapshot model is
shorter than in the AAP model. 
Indeed, being applied to the modeling of channelling, the snapshot model takes  
into account two mechanisms of incoherent scattering that lead to the 
projectile dechanneling.
The fist one is due to the mentioned stochasticity of the scattering angle at a given 
impact parameter resulting from incoherent scattering of the projectile by the 
crystal electrons. The second one is related to the stochastic variation of the 
impact parameter caused by random displacements of the atomic nuclei from 
their equilibrium positions in the crystal due to thermal vibrations.

Only the latter mechanism is present in the AAP model 
implemented in MBN Explorer, hence the longer dechanneling length it predicts:
\begin{equation}
 L_\mathrm{d,M} > L_\mathrm{d,S}. \label{dechanneling_length}
\end{equation}
This fact is interpreted on p. 416 of Ref. \cite{Sushko2013a} in the way that
the snapshot model \textit{``underestimates''} the dechanneling length.
Apparently, the AAP model was adopted in Ref. \cite{Sushko2013a}
as a ``benchmark'',
with respect to which other models have to be evaluated whether they 
``overestimate'' or ``underestimate'' the ``true'' value.

As a matter of fact, the electronic scattering contribution to dechanneling
has been present in the channeling models since the dawn of the field
(see e.g. pp. 36 and 37 of Ref. \cite{Lindhard1965}). Its significance has never been
disputed since then. Therefore, there is absolutely no ground to see the model 
that ignore this effect as a ``benchmark'' or even as a  move 
forward with respect to other models. For example, the algorithm of 
Ref. \cite{Baryshevsky2013a}, which likewise simulates electron channeling form the fist 
principles, also yields a smaller value of the dechanneling length than those 
obtained in Ref. \cite{Sushko2013a}.

Sushko \textit{et al.} refer to the value of the dechanneling length reported in 
Ref. \cite{Backe2008}, which is even larger than the one given by the AAP model, as
to the \textit{``experimental value''}.  
In reality, the dechanneling length was not measured in Ref. \cite{Backe2008} directly.
It was estimated using a diffusion model\footnote{The diffusion model
does not include point-like nuclei of crystal atoms that are present in the snapshot model
as well as in the AAP model. This is the most likely reason for the larger value of the 
dechanneling length.}
within Kitagawa-Ohtsuki approximation \cite{Kitagawa1973}. 
Then the result of the model calculations
was found to be in agreement with a measured dependence of the channeling
radiation intensity on the crystal thickness, which was published in the same paper (Fig. 16 of 
Ref. \cite{Backe2008}). However, the snapshot model was shown to be in agreement 
with the same experimental data as well (Fig. 10 of Ref. \cite{Kostyuk2010}).
In other words, the present experimental data cannot discriminate between 
these models. Hopefully, new experiments at Mainz Microtron (MAMI), in particular those
with crystals of thickness below 10 $\mu$m, will be done and will help one to reduce the 
uncertainty in the electron dechanneling length. 
Similar measurements with a positron beam would be instructive in discriminating between
the snapshot model and the AAP model.

\end{document}